\documentclass[runningheads,openany]{svmult}
\usepackage[T1]{fontenc}
\usepackage{palatino}
\usepackage{euler}
\usepackage{graphicx}  
\usepackage{physprbb}  
\usepackage{bg2002}  
\usepackage{epsfig}

\begin{document}
\setcounter{page}{47}


\let\I\i
\def\i{\mathrm{i}}
\def\e{\mathrm{e}}
\def\d{\mathrm{d}}
\def\half{{\textstyle{1\over2}}}
\def\thalf{{\textstyle{3\over2}}}
\def\h{{\scriptscriptstyle{1\over2}}}
\def\th{{\scriptscriptstyle{3\over2}}}
\def\fh{{\scriptscriptstyle{5\over2}}}
\def\vec#1{\mbox{\boldmath$#1$}}
\def\svec#1{\mbox{{\scriptsize \boldmath$#1$}}}
\def\oN{\overline{N}}
\def\ttimes{{\scriptstyle \times}}
\def\bm#1{{\pmb{\mbox{${#1}$}}}}

\def\CG#1#2#3#4#5#6{C^{#5#6}_{#1#2#3#4}}
\def\threej#1#2#3#4#5#6{\left(\begin{array}{ccc}
    #1&#2&#3\\#4&#5&#6\end{array}\right)}

\title*{%
N-$\Delta$ axial transition form factors
\thanks{Talk delivered by 
B. Golli.
}}

\author{%
B. Golli$^{a,b}$, 
L. Amoreira$^{c,e}$,
M. Fiolhais$^{d,e}$, and
S. \v{S}irca$^{f,b}$
}\institute{%
$^a${Faculty of Education,
              University of Ljubljana,
              1000 Ljubljana, Slovenia}\\
$^b${J.~Stefan Institute, 
              1000 Ljubljana, Slovenia}\\
$^c${Department of Physics,
                  University of Beira Interior,
                  6201-001 Covilh\~a, Portugal}\\
$^d${Department of Physics,
                  University of Coimbra,
                  3004-516 Coimbra, Portugal}\\
$^e${Centre for Computational Physics, University of Coimbra, 
                    3004-516 Coimbra, Portugal}
$^f${Faculty of Mathematics and Physics,
              University of Ljubljana,
              1000 Ljubljana, Slovenia}
}

\authorrunning{B. Golli, L. Amoreira, M. Fiolhais, and S. \v{S}irca}
\titlerunning{N-$\Delta$ axial transition form factors} 

\maketitle

\begin{abstract}
We review some basic properties of the N-$\Delta$ transition 
axial amplitudes and relate them
to the strong $\pi\mathrm{N}\Delta$ form-factor.
In models with the pion cloud we derive a set of constraints 
on the pion wave function which guaranty the correct behaviour 
of the amplitudes in the vicinity of the pion pole.
Corrections due to the spurious center-of-mass motion are 
calculated to the leading order in the inverse baryon mass. 
We give explicit expressions for the amplitudes in 
the Cloudy Bag Model and show that they rather strongly
underestimate the experimental values.
\end{abstract}

\section{Introduction}

The weak N-$\Delta$ transition amplitudes yield important
information about the structure of the nucleon and the $\Delta$, 
and in particular about the role of chiral mesons since they 
explicitly enter in the expression for the axial part of 
the weak current.
There exist only very few calculations in quark models
\cite{bg:Mukh,bg:Hemmert} yet none of them includes the 
mesonic degrees of freedom.
This can be traced back to the difficulty of incorporating 
consistently the pion field which is necessary to describe 
the correct low-$Q^2$ behaviour of the amplitudes.
Obviously, this can be done only in the models that 
properly incorporate the chiral symmetry.

The aim of this work is to study the axial amplitudes of the 
N-$\Delta$ transition in models with quarks and chiral mesons.
In Sec.~2 we introduce expressions for the axial helicity 
amplitudes and relate them to the experimentally measured 
quantities, $C^A_i, i=3,6$, the so called Adler form-factors.
We derive the analog of the Goldberger-Treiman relation
that relates the leading axial form factor, $C_5^A$, 
to the strong $\pi\mathrm{N}\Delta$ coupling constant.
In Sec.~3 we calculate the amplitudes in a simple
isobar model that includes the pion.
In Sec.~4 we study some general properties of the axial
amplitudes in quark models that include the pion and
possibly also its chiral partner, the $\sigma$-meson.
We derive a set of constraints on the pion field and
show that in models that satisfy these constraints
the pion pole appears only in the $C_6^A$ form-factor.
Furthermore, if the meson self-interaction is absent
in the model, i.e. if the pion interacts only with 
quarks, the pion contributes  solely to the $C_6^A$ 
form-factor while the $C_4^A$ and$C_5^A$ form-factors 
pick up only the contribution from quarks.
In most quark models the nucleon and the $\Delta$ are 
calculated as localized states while the expressions for 
the amplitudes require states with good linear momenta.
In Sec.~5 we use the wave packet formalism to derive 
corrections to the amplitudes calculated between localized 
states and show that the approximations  are valid for 
momenta that are small compared to typical baryon masses.
In Sec.~6 we give explicit expressions for the axial 
as well as the strong form-factors in the Cloudy Bag Model 
(CBM) and make a simple estimate of their strengths.

The calculation of the form-factors in the CBM as well as 
in the linear $\sigma$-model that includes besides the pion
also the $\sigma$-meson is presented and compared
to the experimentally measure form-factors in 
\cite{bg:letter,bg:FB18} and in the contribution of 
Simon \v{S}irca~\cite{bg:miniBled} to these Proceedings.

\section{Same basic properties of transition amplitudes}

\subsection{Definition of the helicity amplitudes}

The  weak transition amplitudes are defined as the matrix 
elements of the weak interaction Hamiltonian
\begin{equation}
M =  \langle\Delta|H|N,W\rangle = 
   W_{a\mu}^{(-)}\langle\Delta|V^{a\mu}-A^{a\mu}|N\rangle
\label{Hweak}
\end{equation}
where $a$ is the isospin index.
For simplicity we shall assume $a=0$ and will not write 
it explicitly. 
For the axial part alone we have:
\begin{equation}
 M^A  =  \sqrt{4\pi\alpha_W\over2K_0}\, 
    \sum_\lambda e_{\mu\lambda}\langle\Delta|A^\mu|N\rangle
=
   \sqrt{4\pi\alpha_W\over2K_0}\, 
    \left[\langle\Delta|A^0|N\rangle
         - \sum_\lambda \vec{\varepsilon}_\lambda\cdot 
           \langle\Delta|\vec{A}|N\rangle\right]\,,
\label{Maxial}
\end{equation}
where 
\begin{equation}
 K_0 = {M_\Delta^2 - M_N^2 \over 2 M_\Delta}
\quad\mbox{and}\quad
4\pi\alpha_W = {4\pi\alpha\over\sin^2\theta_W} \approx 0.443\;.
\label{alphaW}
\end{equation}
The 4-momentum of the incident weak boson (W) is
\begin{equation}
 k^\mu =(k_0,0,0,k)\;,
\quad
 k_0 = {M_\Delta^2 - M_N^2 -Q^2 \over 2 M_\Delta}\;,
\quad
 k   = \sqrt{k_0^2 + Q^2}\;.
\label{kin}
\end{equation}

The helicity amplitudes are defined as
\begin{eqnarray}
  \tilde{S}^A &=& -\langle \Delta^+(p'),s_\Delta=\half|A^0_0(0)
                  |N^+(p)s_N=\half\rangle\,,
\label{defS}\\
  \tilde{A}^A_\th &=& -\langle \Delta^+(p'),s_\Delta=\thalf|
                   \vec{\varepsilon}_+\cdot\vec{A}(0)
                  |N^+(p)s_N=\half\rangle\,,
\label{defA32}\\
  \tilde{A}^A_\h &=& -\langle \Delta^+(p'),s_\Delta=\half|
                   \vec{\varepsilon}_+\cdot\vec{A}(0)
                  |N^+(p)s_N=-\half\rangle\,,
\label{defA12}\\
  \tilde{L}^A &=& -\langle \Delta^+(p'),s_\Delta=\half|
                   \vec{\varepsilon}_0\cdot\vec{A}(0)
                  |N^+(p)s_N=\half\rangle\,.
\label{defL}
\end{eqnarray}


\subsection{The Adler form-factors}

Experimentalists measure the so called Adler  form-factors
defined as \cite{bg:llewellyn72}:
\begin{eqnarray}
\langle \Delta^+(p')|A_{\alpha(a=0)}|\mathrm{N}^+(p)\rangle
 &=& 
   \bar{u}_{\Delta\alpha}\,{C^\mathrm{A}_4(Q^2)\over M_\mathrm{N}^2}\,
     {p'}_\mu q^\mu u_\mathrm{N}
   - \bar{u}_{\Delta\mu}\,{C^\mathrm{A}_4(Q^2)\over M_\mathrm{N}^2}\,
     {p'}_\alpha q^\mu u_\mathrm{N}\nonumber\\
&&\kern-33mm
   + \bar{u}_{\Delta\alpha}\,C^\mathrm{A}_5(Q^2)\,u_\mathrm{N}
   + \bar{u}_{\Delta\mu}\,{C^\mathrm{A}_6(Q^2)\over M_\mathrm{N}^2}\,
     q^\mu q_\alpha u_\mathrm{N}
   + \bar{u}_{\Delta\alpha}\,{C^\mathrm{A}_3(Q^2)\over M_\mathrm{N}}\,
     \gamma_\mu q^\mu  u_\mathrm{N}\;,
\label{Aexp}
\end{eqnarray}
where ${p'}_\mu=(M_\Delta;0,0,0)$ and $q^\mu=(\omega;0,0,k)$,
and ${u}_{\Delta\alpha}$ is the Rarita-Schwinger spinor:
\begin{equation}
  u_\alpha(p,s_\Delta) =\sum_{\lambda',s}
         \CG{1}{\lambda'}{\h}{s}{\th}{s_\Delta} 
         e_{\alpha\lambda'}(p) u(p,s)\,.
\label{RSspinor}
\end{equation}
Here
\begin{equation}
  e^\mu_\lambda(p) = \left[
   {\vec{\varepsilon}_\lambda\cdot\vec{p}\over M_\Delta}\;, \>
   \vec{\varepsilon}_\lambda + 
       {\vec{p}(\vec{\varepsilon}_\lambda\cdot\vec{p})\over
         M_\Delta(p_0+M_\Delta)}\right]\;,
\label{emulambda}
\end{equation}
and $u(p,s)$ is the usual bispinor for a spin $\half$ 
particle.
For the $\Delta$ at rest it has a simple form 
(e.g. \cite{bg:Weise}, 414):
\begin{equation}
  e^\mu_\lambda= (0\,,\,\vec{\varepsilon}_\lambda)\;,
\qquad
  u(p,s) = {1\choose0}\chi_{\h s}\;,
\label{RS0}
\end{equation}
where $\vec{\varepsilon}_\lambda$ are the polarization vectors.
The form-factor $C^A_3$ is small; in models with $s$-wave quarks 
and $p$-wave pions it is even identically  0;
we shall therefore assume $C^A_3=0$ in the further derivations.
  
The helicity amplitudes can now be easily related
to the form factors.
For $\alpha=0$ the evaluation is straightforward, while
for $\alpha\ne0$ we multiply (\ref{Aexp}) by $e^\alpha_\lambda$
and use the following relations:
\begin{equation}
 e^\alpha_\lambda \bar{u}_\alpha(p,s_\Delta) u_N 
= \vec{\varepsilon}_\lambda \sum_{\lambda',s}
         \CG{1}{\lambda'}{\h}{s}{\th}{s_\Delta}
      (-\vec{\varepsilon}^*_{\lambda'})\bar{u}(p,s)u_N
= - \CG{1}{\lambda}{\h}{s_N}{\th}{s_\Delta}\;,
\label{rs1}
\end{equation}
\begin{equation}
 e^\alpha_\lambda q_\alpha = -k\delta_{\lambda,0}\;,
\qquad
 \bar{u}_\mu(p,s_\Delta) q^\mu u_N = 
       -k \CG{1}{\lambda}{\h}{s_N}{\th}{s_\Delta}\;.
\label{rs2}
\end{equation}
We obtain
\begin{eqnarray}
\tilde{S}^A
  &=& 
          -\left[k\,{C^A_4\over M_N^2}\,M_\Delta
               - \omega k {C^A_6\over M_N^2}\right]\sqrt{2\over3}\,,
\label{SAexp}\\
\tilde{A}^A_\th
  & = &
         -\left[{C^A_4\over M_N^2}\,\omega M_\Delta + C^A_5\right]
    =      \sqrt{3}\,\tilde{A}^A_\h\;,
\label{AAexp}\\
\tilde{L}^A
  & = &
         -\left[{C^A_4\over M_N^2}\,\omega M_\Delta + C^A_5
           -{k^2\over M_N^2} C^A_6\right]\sqrt{2\over3}\;.
\label{LAexp}
\end{eqnarray}
The Adler form-factors read
\begin{eqnarray}
C_6^A  & = & {M_N^2\over k^2}\,
    \left[-\tilde{A}^A_\th + \sqrt{3\over2}\tilde{L}^A\right]\,,
\label{C6}\\
C_5^A  & = &
  -\sqrt{3\over2}\left(\tilde{L}^A - {k_0\over k}\,\tilde{S}^A\right)
  - {k_0^2-k^2\over M_N^2}\,C_6^A\,,        
\label{C5}\\
C_4^A  & = &
 {M_N^2\over kM_\Delta}\left[-\sqrt{3\over2}\,\tilde{S}^A
 + {k_0k\over M_N^2}\,C_6^A  \right]\,.
\label{C4}
\end{eqnarray}

\subsection{The off-diagonal Goldberger-Treiman relation}

Let us compute the divergence of the axial current
between the $\Delta$ and $N$ (\ref{Aexp}).
Using (\ref{rs2}) we get ($q^2\equiv-Q^2$):
\begin{equation}
 \langle \Delta^+(P)|\partial^\alpha A_\alpha|N^+(p)\rangle 
  = \i k\left[C^A_5(q^2) + {C^A_6(q^2)\over M_N^2}\,q^2\right]
      \CG10\h\h\th\h\;.
\label{dAexp} 
\end{equation}
In the chiral limit the divergence has to vanish.
From the above expression we would conclude that $C^A_5(q^2)=0$
which is experimentally not the case.
Hence  $C^A_6(q^2)$ should have a pole at $q^2=0$ such that
\begin{equation}
  C^A_6(q^2) = - {M_N^2 C^A_5(q^2)\over q^2}\,.
\label{C6toC5} 
\end{equation}
As in the nucleon case, we relate this term to the term 
in the axial current that is responsible for the pion decay:
${A^\alpha_a}_\mathrm{pole}(x) = f_\pi\partial^\alpha\pi_a(x)$.
We can therefore identify the $C_6^A$-term 
in (\ref{Aexp}) with:
\begin{equation}
 \bar{u}_{\Delta\mu}\,{C^A_6(q^2)\over M_N^2}\,q^\mu q_\alpha u_N
=
 \i q_\alpha f_\pi\langle \Delta^+(P)|\pi_0(0)|N^+(p)\rangle\,. 
\label{AC6} 
\end{equation}
Indeed, the pion propagator behaves as $q^{-2}$ in the chiral limit.

In the real world the pion mass is finite and we write the pion
field as
\begin{equation}
  \langle \Delta^+(P)|\pi_0(0)|N^+(p)\rangle
 = \i {G_{\pi\mathrm{N}\Delta}(q^2)\over 2M_N}\,
 {\bar{u}_{\Delta\mu}\,q^\mu u_N\over -q^2 + m_\pi^2}\, 
 \sqrt{2\over3}\;. 
\label{piDN} 
\end{equation}
while the vanishing of (\ref{dAexp}) is replaced by PCAC:
\begin{equation}
 \langle \Delta^+(P)|\partial^\alpha A_{\alpha\,a}|N^+(p)\rangle 
  = -m_\pi^2\,f_\pi\langle \Delta^+(P)|\pi_a(0)|N^+(p)\rangle\,.
\label{dA} 
\end{equation}
Replacing the LHS of (\ref{dA}) by (\ref{dAexp}) and using
(\ref{AC6}) and (\ref{piDN}) we find
\begin{eqnarray}
\i q^\alpha\bar{u}_{\Delta\alpha} u_N\left[
        C^A_5(q^2) + f_\pi{G_{\pi\mathrm{N}\Delta}(q^2)\over 2M_N}
       {q^2\over -q^2 + m_\pi^2}\, \sqrt{2\over3}   \right]
  &=&
\nonumber\\
&&\kern-40mm
 \i q^\alpha\bar{u}_{\Delta\alpha} u_N
   {G_{\pi\mathrm{N}\Delta}(q^2)\over2M_N}
     {m_\pi^2\,f_\pi\over -q^2+m_\pi^2}
  \, \sqrt{2\over3}\;. 
\label{dAexp1} 
\end{eqnarray}
We finally obtain
\begin{equation}
    C^A_5(q^2) = f_\pi\,{G_{\pi\mathrm{N}\Delta}(q^2)\over 2M_N}\, 
  \sqrt{2\over3}\;,
\label{GT0} 
\end{equation}
the {\em off-diagonal Goldberger-Treiman relation\/}, 
which -- strictly speaking -- holds only in the limit 
$q^2\rightarrow m_\pi^2$.
Assuming a smooth behaviour of the amplitudes for $q^2$
in the vicinity of $m_\pi^2$ we can expect (\ref{GT0})
to remain valid for sufficiently small $q^2$ in the
experimentally accessible range.


\section{The axial current in a simple isobar model with pions}

The aim of this section is to derive the amplitudes in a simple 
model in order to study the contribution of pions to the amplitudes 
and to analyze the qualitative behaviour of the amplitudes. 
The derivation in this section is based on the standard
derivation of the diagonal Goldberger-Treiman relation and PCAC 
(see e.g. \cite{bg:Weise}).

We investigate the axial hadronic current in a model with two 
structureless fermion fields, the nucleon and the $\Delta$,
and the pion field.
Since we are interested here only in the nucleon-$\Delta$
transition we shall write down explicitly only the pertinent 
parts of the Lagrangian and of the hadron current.
The nucleon and the $\Delta$ (at rest) satisfy the Dirac equation
\begin{equation}
  (\i\gamma_\mu\partial^\mu -E_N)\psi_N = 0\;,
\qquad
  (\i\gamma_\mu\partial^\mu -M_\Delta)\psi_\Delta = 0\;.
\label{Dirac}
\end{equation}
We assume the following form of the $\pi\mathrm{N}\Delta$ interaction
\begin{equation}
  \mathcal{L}_{\pi\mathrm{N}\Delta} = -\i G_{\pi\mathrm{N}\Delta}
                     \bar\psi_\Delta\gamma_5T_a\psi_N\pi_a\;,
\label{LNDpi}
\end{equation}
where we introduce the transition operator $\pol{T}$ 
(and $\vec{\Sigma}$) by
\begin{equation}
  \langle \thalf t_\Delta|T_a|\half t_N\rangle 
     = \CG{1}{a}{\h}{t_N}{\th}{t_\Delta}\;,
\qquad
  \langle \thalf s_\Delta|\Sigma_\lambda|\half s_N\rangle 
     = \CG{1}{\lambda}{\h}{s_N}{\th}{s_\Delta}\;.
\label{TSigma}
\end{equation}
(Note that $\gamma^\mu$ has a more complicated structure:
\begin{equation}
  \vec{\gamma} = \left|\begin{array}{cc} 0 & \vec{S} \\
                                  -\vec{S} & 0 \end{array}\right|\,,
\label{gammaT}
\end{equation}
where the generalized Pauli matrices $\vec{S}$ act in the space 
spanned by the $S=\half$ and $S=\thalf$ subspaces:
\begin{equation}
  \vec{S} = \left|\begin{array}{cc} \vec{\sigma} & \vec{\Sigma} \\
                   \vec{\Sigma}^\dagger & \vec{\sigma}_{\Delta\Delta} 
              \end{array}\right|\;.
\label{spinT}
\end{equation}
The generalized isospin is introduce in the same way.)

The nucleon bispinor can be written as
\begin{equation}
  u_N(p) = \sqrt{E_N+M_N\over2M_N}
            {1\choose{\svec{\Sigma}\cdot\svec{p}\over E_N+M_N}} 
	    \chi_{\h s_N}\xi_{\h t_N}
            \approx {1\choose{\svec{\Sigma}\cdot\svec{p}\over 2M_N}} 
	    \chi_{\h s_N}\xi_{\h t_N}\;,
\label{uN}
\end{equation}
with $\chi$ and $\xi$ describing respectively the spin and isospin
part of the bispinor, and
\begin{equation}
  p^\mu = (E_N,\vec{p})\;,
\qquad
  E_N=\sqrt{M_N^2 + \vec{p}^2} \approx M_N\;.
\label{pDpN}
\end{equation}
We assume that $\Delta$ is at rest, ${p'}^\mu = (M_\Delta;0,0,0)$,
hence 
\begin{equation}
  u_\Delta(p') = {1\choose0}\chi_{\th s_\Delta}\xi_{\th t_\Delta}\;.
\label{uDelta}
\end{equation}

In the model, the transition part of the axial current takes the form:
\begin{equation}
  A^\mu_a = 
     g_A^\Delta \bar{\psi}_\Delta\gamma^\mu\gamma_5\,\half T_a\psi_N
        + f_\pi\partial^\mu\pi_a   \;.
\label{AxialT}
\end{equation}
Using the Dirac equations (\ref{Dirac}) and the Klein-Gordon equation
for the (pertinent part of the) pion field:
\begin{equation}
 \left(\partial_\mu\partial^\mu + m_\pi^2\right)\pi_a
   =-\i G_{\pi\mathrm{N}\Delta}\bar\psi_\Delta\gamma_5T_a\psi_N
\label{KG}
\end{equation}
 we immediately obtain
\begin{equation}
  \partial_\mu A^\mu_a = \i g_A^\Delta\,\half(M_\Delta + M_N)
   \bar{\psi}_\Delta\gamma_5\,T_a\psi_N          
   -\i f_\pi G_{\pi\mathrm{N}\Delta}\bar\psi_\Delta\gamma_5T_a\psi_N
    -f_\pi m_\pi^2\,\pi_a\;. 
\label{dAxialT}
\end{equation}
In the limit $m_\pi\rightarrow0$ the current is conserved provided
\begin{equation}
  \half(M_\Delta + M_N)g_A^\Delta = f_\pi G_{\pi\mathrm{N}\Delta} 
\label{GT}
\end{equation}
which is the {\em off-diagonal Goldberger-Treiman relation\/}
(\ref{GT0}).
The constant $g_A^\Delta$ is related to the experimentally measured
$C_5^A(0)$ by
\begin{equation}
  g_A^\Delta = {2M_N\over M_\Delta + M_N}\,\sqrt{6}\, C_5^A(0)\;,
\qquad
  C_5^A(0) = 1.22\pm0.06\;.
\label{gAD}
\end{equation}

We now evaluate the matrix elements of the transition axial current.
In this case the solution of (\ref{KG}) is
\begin{equation}
  \langle \Delta(p')|\pi_a(\omega,\vec{k})|N(p)\rangle = 
    -\i \; {G_{\pi\mathrm{N}\Delta}\over2M_N}\;
      {\langle \Delta|\left(-\vec{\Sigma}\cdot\vec{k}\right)\, 
         T_a|N\rangle
          \over (-\omega^2 +\vec{k}^2 + m_\pi^2)}
\label{pia}
\end{equation}
with $\omega = M_\Delta - M_N$, $\vec{k} = -\vec{p}$.
For the time-like component of the current we get
\begin{eqnarray}
\langle \Delta(p')|A^0_a(0)|N(p)\rangle
  &=& -k\,{g_A^\Delta\over2M_N}
    \langle \Delta|\Sigma_0\,\half T_a|N\rangle 
    + \i\omega f_\pi\langle \Delta(p')|\pi_a|N(p)\rangle
\nonumber\\
  &=& - \left[{g_A^\Delta\,k\over4M_N} + 
      {f_\pi G_{\pi\mathrm{N}\Delta}\over2M_N}\;
              {\omega k\over(-q^2+m_\pi^2)}
      \right]\langle \Delta|\Sigma_0\,T_a|N\rangle\,.
\label{Aa0}
\end{eqnarray}
The spatial part is
\begin{eqnarray}
\langle \Delta(p')|\vec{A}_a(0)|N(p)\rangle
  &=& 
    g_A^\Delta
    \langle \Delta|\vec{\Sigma}\,\half T_a|N\rangle 
    + \i \vec{k} f_\pi \langle \Delta(p')|\pi_a|N(p)\rangle
\nonumber\\
  &=& 
   \half\, g_A^\Delta\langle \Delta|\vec{\Sigma}\, T_a|N\rangle
      -  {f_\pi G_{\pi\mathrm{N}\Delta}\over2M_N}\;
              {\vec{k}\over(-q^2+m_\pi^2)}\;
      \langle \Delta|(\vec{\Sigma}\cdot\vec{k})\,T_a|N\rangle\;.
\nonumber\\
\label{Aavec}
\end{eqnarray}

The helicity amplitudes introduced in the first section 
(for 4-vector momentum transfer 
$q^\mu={p'}^\mu - p^\mu = (\omega;0,0,k)$)
are now expressed as
\begin{eqnarray}
  \tilde{S}^A 
  &=& \left[ k \, {g_A^\Delta\over4M_N} + 
      k\,{f_\pi G_{\pi\mathrm{N}\Delta}\over2M_N}\;
              {\omega\over(-q^2+m_\pi^2)}
      \right]\sqrt{2\over3}^2\,,
\label{SAt}\\
  \tilde{A}^A_\th &=& -\half\, g_A^\Delta\sqrt{2\over3} 
                   = \sqrt{3}\tilde{A}^A_\h\;,
\label{A3t}\\
  \tilde{L}^A
   &=& \left[-\half\, g_A^\Delta +
        {f_\pi G_{\pi\mathrm{N}\Delta}\over2M_N}\;
              {k^2\over(-q^2+m_\pi^2)} \right]\sqrt{2\over3}^2\,.
\label{LAt}
\end{eqnarray}
Using (\ref{GT}) we are now able to explicitly check that PCAC
holds in the model:
\begin{eqnarray}
 \langle \Delta^+(p')|
        \partial_\mu A^\mu_{a=0}|N^+(p)\rangle
   &=& -\i \left(\omega \tilde{S}^A - k\tilde{L}^A\right)
\nonumber\\
  &=& - m_\pi^2 f_\pi \langle \Delta^+(p')|\pi_0|N^+(p)\rangle\;.
\label{PCAC}
\end{eqnarray}
In this model we can express the Adler form-factors solely 
in terms of either $g_A^\Delta$ or $G_{\pi\mathrm{N}\Delta}$:
\begin{eqnarray}
  C_6^A  & = & {1\over\sqrt6}\,
         f_\pi M_N\,{G_{\pi\mathrm{N}\Delta}\over -q^2+m_\pi^2}\,,
\label{C6p}\\
C_5^A  & = &  
   {1\over\sqrt6}\, {M_\Delta+M_N\over2M_N}\, g_A^\Delta
    = \sqrt{2\over3}\,{f_\pi G_{\pi\mathrm{N}\Delta}\over2M_N}\,, 
\label{C5p}\\
C_4^A  & = & -{1\over\sqrt6}\, {M_N\over2M_\Delta}\, g_A^\Delta
         = - {M_N^2\over M_\Delta(M_\Delta+M_N)}\,C^A_5
         \approx -0.33\,C^A_5\;. 
\label{C4p}
\end{eqnarray}
The relations derived above show that only $C^A_6$ exhibits 
the pole behavior while in the other two amplitudes the pole 
behavior cancels out and the result is the same as 
if we used only the fermion part of the axial current.
In the next section we shall see that this property holds in 
a vast class of models that fulfill certain virial relations.


\section{Helicity amplitudes in models with the pion cloud}

We investigate  quark models that include the pion and
possibly also its chiral partner, the $\sigma$-meson.
The part of the Hamiltonian that involves pions
can be written in the following form:
\begin{equation}
   H_\pi = \int\d \vec{r}\left\{\half\left[\pol{P}_\pi^2 + 
        (\nabla^2 + m_\pi^2)\pol{\pi}^2\right]
        + U(\sigma,\pol{\pi}) + 
         \sum_t j_t\pi_t\right\}\,.
\label{Hchi}
\end{equation}
Here $j_t$ represents the quark pseudoscalar-isovector 
source term, $t$ is the third component of the isospin,
and $U(\sigma,\pol{\pi})$ a possible meson self-interaction
term (such as the Mexican hat potential of the linear 
$\sigma$-model).
Let $|N\rangle$ and  $|\Delta\rangle$ be the ground
state and the excited state describing the $\Delta$
with $H|N\rangle = E_N|N\rangle$ and 
$H|\Delta\rangle=E_\Delta|\Delta\rangle$, then we
can write the following virial theorems (relations):
\begin{eqnarray}
 \langle N|[H,\pol{P}_\pi]|N\rangle
  &=& 
 \langle N|H \pol{P}_\pi - \pol{P}_\pi H|N\rangle
  = 0\,,
\label{virN}\\
 \langle \Delta|[H,\pol{P}_\pi]|\Delta\rangle
  & = & 0\,,
\label{virD}\\
 \langle \Delta|[H,\pol{P}_\pi]|N\rangle
  & = & 
  (E_\Delta - E_N) \langle \Delta|\pol{P}_\pi|N\rangle
  = \i(E_\Delta - E_N)^2 \langle \Delta|\pol{\pi}|N\rangle\,.
\label{virDN}
\end{eqnarray}
We have used $\pol{P}_\pi=\i[H,\pol{\pi}]$ in the last line.
We call (\ref{virDN}) the {\em off-diagonal virial relation 
(theorem)\/}.
(Note that there is no off-diagonal relation of this type for
the $\sigma$-field because it is scalar-isoscalar and the matrix
elements vanish identically.)

We now evaluate the commutators on the LHS using (\ref{Hchi}):
\begin{eqnarray}
 (-\Delta+m_\pi^2)\langle N|\pi_t(\vec{r})|N\rangle
  &=& 
  -(-1)^t \langle N|J_{-t}(\vec{r})|N\rangle\,,
\label{virN-KG}\\
 (-\Delta+m_\pi^2)\langle\Delta|\pi_t(\vec{r})|\Delta\rangle
  &=& 
  -(-1)^t \langle\Delta|J_{-t}(\vec{r})|\Delta\rangle\,,
\label{virD-KG}\\
 (-\Delta+m_\pi^2-\omega_*^2)\langle\Delta|\pi_t(\vec{r})|N\rangle
  & = & 
  -(-1)^t \langle\Delta|J_{-t}(\vec{r})|N\rangle\,.
\label{virDN-KG}
\end{eqnarray}
We have defined $\omega_*=(E_\Delta - E_N)$ and
\begin{equation}
  J_t(\vec{r}) = j_t(\vec{r}) + (-1)^t\,
   {\partial U(\sigma,\pol{\pi})\over\partial\pi_{-t}(\vec{r})}\,,
\label{Jt}
\end{equation}
and used 
\begin{equation}
   [\pi_{t'}(\vec{r}'), P_{\pi\,,t}(\vec{r})]
           =\i (-1)^t\delta_{t,-t'}\delta(\vec{r}'-\vec{r})\,.
\label{compiP}
\end{equation}

These relations hold for the exact solutions; in an
approximate computational scheme we can use these
relations as constraints on the approximate states.

We now show an important property of the axial transition
amplitudes which holds for the states that satisfy the above
virial relations.
Let us split the axial current into two parts:
\begin{eqnarray}
  \pol{A}^\alpha
  &=& 
  \pol{A}^\alpha_{np}+ \pol{A}^\alpha_{pole}\;,
\label{Asplit}\\
\pol{A}^\alpha_{np}
  & = &
    \bar{\psi}\gamma^\alpha\gamma_5\,\half\pol{\tau}\psi 
  + (\sigma-f_\pi)\partial^\alpha{\pol{\pi}} 
    - {\pol{\pi}}\partial^\alpha{\sigma}\,,
\label{Anp}\\
\pol{A}^\alpha_{pole}
   & = & f_\pi\partial^\alpha \pol{\pi}\,.
\label{Apole}
\end{eqnarray}
We can now relate the non-pole contribution (\ref{Anp})
to the first term in (\ref{AxialT}) and (obviously)
the pole contribution to the second term in (\ref{AxialT}).
Since the off-diagonal virial relation (\ref{virDN-KG})
coincides with (\ref{pia}), the evaluation is similar to
the derivation presented in the previous section.
The pole term (\ref{Apole}) contributes only to the
longitudinal and the scalar amplitude, hence:
\begin{eqnarray}
 C^A_{6\,(pole)}
  &=& 
  -\i f_\pi\,{M_N^2\over k}\,\sqrt{3\over2}\,
  \langle\Delta^+_{s_\Delta=\h}|\pi_0(0)|N^+_{s_n=\h}\rangle\,,  
\label{C6pole}\\
 C^A_{5\,(pole)} & = & 0\,,
\nonumber\\
 C^A_{4\,(pole)} & = & 0\,.
\nonumber
\end{eqnarray}


\section{Calculation of form-factors between localized states}

The amplitudes (\ref{defS})-(\ref{defL}) are defined between
states with good 4-momenta $p'$ and $p$ respectively while
in the model calculations localized states are used.  We can use
such states in our calculation of amplitudes by interpreting them
as wave packets of states with good linear momenta:
\begin{equation}
 |B(\vec{r})\rangle = 
        \int\d\vec{p}\, \varphi(\vec{p})\,\e^{\i\svec{p}\cdot\svec{r}} 
         |B(\vec{p})\rangle \,.
\label{Blocal}
\end{equation}
The spin-momentum dependence of $|B(\vec{p})\rangle$ is expressed
by the bispinor
\begin{equation}
   u_B(\vec{p}) = \sqrt{E+M\over2M}
   {1\choose{\svec{\sigma}\cdot\svec{p}\over E+M}}\chi_\mathrm{spin}\,.
\label{uBp}
\end{equation}
Requiring (\ref{uBp}) is normalized,   
$\langle B(\vec{p})|B(\vec{p})\rangle=1$, we have
\begin{equation}
   \int\d\vec{r}\langle B(\vec{r})|B(\vec{r})\rangle =
   (2\pi)^3\int\d\vec{p}\,|\varphi(\vec{p})|^2 = 1\;.     
\label{normBr}
\end{equation}

We now relate matrix elements between localized states 
to matrix elements between states with good momenta.
We start by a matrix element between localized states:
\begin{eqnarray}
 \int\d\vec{r}\, e^{\i\svec{k}\cdot\svec{r}}
     \langle\Delta|M(\vec{r})|N\rangle
  &=& \int\d\vec{r}\int\d\vec{p}'\int\d\vec{p}\,
      e^{\i(\svec{k}-\svec{p}'+\svec{p})\cdot\svec{r}}
     \langle\Delta(\vec{p}')| M(\vec{r})|N(\vec{p})\rangle
\nonumber\\
  &&\times   \varphi_\Delta^*(\vec{p}')\varphi_N(\vec{p}).
\label{Mloc}
\end{eqnarray}
Since the matrix element 
$\langle\Delta(\vec{p}')| M(\vec{r})|N(\vec{p})\rangle$
does not depend on $\vec{r}$ (all $\vec{r}$-depen\-dence
is contained in the exponential) we can substitute
it by its value at $\vec{r}=0$.
We then carry out the $\vec{r}$ integration yielding
$\delta(\vec{p}-\vec{p}'+\vec{k})$, and the above matrix 
element reads:
\begin{equation}
 \int\d\vec{r}\, e^{\i\svec{k}\cdot\svec{r}}
     \langle\Delta|M(\vec{r})|N\rangle
  =   (2\pi)^3\int\d\vec{p}\,
  \langle\Delta(\vec{p}+\vec{k})| M(0)|N(\vec{p})\rangle\;
  \varphi_\Delta^*(\vec{p}+\vec{k})\varphi_N(\vec{p})\;.
\label{Mloc1}
\end{equation}
From the parameterization of the axial current (\ref{Aexp})
we can read off the $p'$ and $p$ dependence 
and plug it into (\ref{Mloc1}).
We neglect terms of the order $p^2/M^2$, e.g. the last 
term in the expression
(\ref{emulambda}) for $e^\mu_\lambda(p)$.
We find:
\begin{equation}
\bar{u}_\alpha(p',s_\Delta=\half)q^\alpha u_N(s=\half)
 = \left[{M_\Delta-M_N\over M_\Delta} {p'}_3-k\right]\sqrt{2\over3}
\label{uDquN}
\end{equation}
and
\begin{equation}
\bar{u}_0(p',s_\Delta=\half) u_N(s=\half)
 = {{p'}_3\over M_\Delta}\sqrt{2\over3}\;.
\label{uD0uN}
\end{equation}
We can carry out the integration over $\vec{p}$
since $C_i(q^2)$ do not depend on $\vec{p}$.
We assume $\varphi_\Delta(\vec{p}) \approx \varphi_N(\vec{p})
\equiv \Pi_{i=1}^3\varphi(p_i)$. 
A typical integral gives:
\begin{eqnarray}
  (2\pi)^3\int\d\vec{p}\, p_3\,
  \varphi(\vec{p}+\vec{k})\varphi(\vec{p}) 
&=& 
   2\pi\int\d p_3\, p_3 \varphi(p_3+k)\varphi(p_3)
\nonumber\\
&=&   
  2\pi\int\d q\, (q-\half k)\,\varphi(q+\half k)\varphi(q-\half k)
\nonumber\\
&=& 
  -\half k\left[1-\half k^2 \int\d q\,\varphi'(q)^2+\ldots\right]
\nonumber\\
&\approx& 
 -\half k\left[1-\half k^2\langle z_\mathrm{c.m.}^2\rangle\right]\;,
\label{p3int}
\end{eqnarray}
where we have taken into account that $\varphi$ are normalized
and used the relation ($\tilde{\varphi}(z)$ is the Fourier 
transform of $\varphi(q)$):
\begin{equation}
  \int\d q\, \varphi'(q)^2 = \int\d z\, z^2\tilde{\varphi}(z)^2
   =  \langle z^2\rangle\;.
\label{intz}
\end{equation}
(Integrating ${p'}_3$ we would get $\half k$.)
Here $\langle z_\mathrm{c.m.}^2\rangle= 
{1\over3}\langle r_\mathrm{c.m.}^2\rangle$ is a typical spread  
of the wave packet describing the center-of-mass motion of 
the localized state and is of the order of the inverse baryon mass.
Clearly, in this approximation it is not meaningful to 
calculate the form-factor to very high $k$.
We finally obtain (neglecting terms of the order $k^2/M^2$):
\begin{eqnarray}
\tilde{S}^A 
  &=& 
    -\left[k\,{M_\Delta\over M_N^2}\,C^A_4 
                + {k\over2M_\Delta}\,C^A_5
   - {\omega k\over M_N^2}\,{M_\Delta+M_N\over2M_\Delta}\,C^A_6\right]
               \sqrt{2\over3}\,,
\label{SAwp}\\
\tilde{A}^A_\th = 
  & = &
         -\left[\omega\,{M_\Delta\over M_N^2}\,C^A_4 + C^A_5\right]
    =      \sqrt{3}\,\tilde{A}^A_\h\,,
\label{AAwp}\\
\tilde{L}^A = 
  & = &
         -\left[\omega\,{M_\Delta\over M_N^2}\,C^A_4 + C^A_5
           -{k^2\over M_N^2}\,{M_\Delta+M_N\over2M_\Delta}\, C^A_6
         \right]\sqrt{2\over3}\,.
\label{LAwp}
\end{eqnarray}
We now express the experimental amplitudes in terms of the
helicity amplitudes as
\begin{eqnarray}
C_6^A  & = & {M_N^2\over k^2}\,
    \left[-\tilde{A}^A_\th + \sqrt{3\over2}\tilde{L}^A\right]\,
    {2M_\Delta\over M_\Delta+M_N}\,,
\label{C6wp}\\
C_5^A  & = &
  -\sqrt{3\over2}\left(\tilde{L}^A - {k_0\over k}\,\tilde{S}^A\right)\,
     {2M_\Delta\over M_\Delta+M_N}
  - {k_0^2-k^2\over M_N^2}\,C_6^A\;,        
\label{C5wp}\\
C_4^A  & = &
 {M_N^2\over kM_\Delta}\left[-\sqrt{3\over2}\,\tilde{S}^A
 + {k_0k\over M_N^2}\,{M_\Delta+M_N\over2M_\Delta}\,C_6^A  \right]
 - {M_N^2\over2M_\Delta^2}\,C_5^A \;.
\label{C4wp}
\end{eqnarray}

The strong form-factor can be treated in the same way.
The general coupling of the pion field to the baryon
is written in the form
\begin{equation}
   H_{B-\pi} = \int\d\vec{r}\, J^\pi_a(\vec{r})\pi_a(\vec{r})\,,
\label{HBpi}
\end{equation}
where $J^\pi_a(\vec{r})$ is the baryon strong pseudoscalar-
isovector current.
The N-$\Delta$ transition matrix element is parameterized as
\begin{equation}
\langle \Delta^+(p')|J^\pi_a(0)|N^+(p)\rangle =
-\i\bar{u}_{\Delta\mu}\,{G_{\pi\mathrm{N}\Delta}(q^2)\over 2M_N}
  \,q^\mu u_N \,,
\end{equation}
where $q=p'-p$.
Using (\ref{uDquN}) we find
\begin{equation}
\langle \Delta^+(p')|J^\pi_a(0)|N^+(p)\rangle =
-\i{G_{\pi\mathrm{N}\Delta}(q^2)\over 2M_N}
\left[{M_\Delta-M_N\over M_\Delta} {p'}_3-k\right]\CG10\h\h\th\h\;.
\end{equation}
We now use of relation (\ref{Mloc1}) as well as (\ref{p3int})
to obtain
\begin{equation}
 {G_{\pi\mathrm{N}\Delta}(q^2)\over 2M_N}
  \,{M_\Delta + M_N\over2M_\Delta}
  = {1\over\i k}  
   \langle\Delta||\int\d\vec{r}\e^{i\svec{k}\cdot\svec{r}}
       J(\vec{r})||N\rangle \,.
\label{defGwp}
\end{equation}


\section{Helicity amplitudes in the Cloudy Bag Model}

The Cloudy Bag Model (CBM) is the simplest example of a quark model 
with the pion cloud that fulfills the virial  constraints 
(\ref{virN})-(\ref{virDN}) provided we take the usual perturbative 
form for the pion profiles \cite{bg:thomas,bg:IJMPh}.
We also take the N-$\Delta$ splitting equal to the
experimental value, $\omega\equiv M_\Delta - E_N$.
Since the pion contribution to the axial current has the form 
of the pole term in (\ref{Apole}), only the quarks contribute
to the $C^\mathrm{A}_5$ and $C^\mathrm{A}_4$ amplitudes.

The helicity amplitudes and the Adler form-factors 
simplify further if we make the usual assumption
of the same quark profiles for the nucleon and the $\Delta$.
In this case the scalar amplitude picks up only the pion
contribution while the quark term  is identically zero.
The transverse amplitude $\tilde{A}^A_\th=\sqrt{3}\,\tilde{A}^A_\h$
has only the quark contribution while 
the longitudinal amplitude has both:
\begin{eqnarray}
\tilde{A}^A_\th(Q^2)
  &=& - {1\over\sqrt{6}}
      \int\d r\,r^2\left[j_0(kr)
      \left(u^2 - {1\over3} v^2\right)
    +{2\over3}\,\,j_2(kr)v^2\right]
	\langle\Delta||\sum\sigma\,\tau||N\rangle ,
\label{Acbm}\\
  \tilde{L}^A(Q^2) 
  &=& 
     -{2\over3}\left\{
     \half\int\d r\,r^2\left[j_0(kr)
      \left(u^2 - {1\over3} v^2\right)
    -{4\over3}\,\,j_2(kr)v^2\right]
\right.
\nonumber\\
  &&\left.  
    -{\omega_\mathrm{MIT}\over\omega_\mathrm{MIT}-1}\,
    {m_\pi\over2f_\pi}\,{j_1(kR)\over kR}\,
   {k^2\over(Q^2+m_\pi^2)}
  \right\}\, 
   \langle\Delta||\sum\sigma\,\tau||N\rangle\,,
\label{Lcbm}\\
\tilde{S}^A(Q^2) 
  &=& 
      {2\over3}\,
    {\omega_\mathrm{MIT}\over\omega_\mathrm{MIT}-1}\,
    {m_\pi\over2f_\pi}\,{j_1(kR)\over kR}\,
    {\omega k\over(Q^2+m_\pi^2)}\,
    \langle\Delta||\sum\sigma\,\tau||N\rangle\,.
\label{Scbm}
\end{eqnarray}
Here $k$ and $Q^2\equiv-q^2$ are related through (\ref{kin}),
$\omega_\mathrm{MIT}=2.04$, and
\begin{eqnarray}
\langle\Delta ||\sum\sigma\tau||N\rangle
&=&\sqrt{Z_N Z_\Delta}
\left\{\rule{0mm}{6mm}\;2\sqrt{2} \right.
\nonumber\\
&&
+ \frac{\sqrt{2}}{27\pi}\;\mathcal{P}
    \int_0^\infty dk\,k^2\,\rho^2(k)
   \left[{25\over\omega_k^2(\omega_k-\omega)}
        + {2\over\omega_k(\omega_k^2-\omega^2)}\right]
\nonumber\\
&& \left.
+\frac{25\sqrt{2}}{27\pi}
    \int_0^\infty dk\,k^2\,\rho^2(k)
 \left[{5\over4\omega_k^3}
     + {1\over\omega_k^2(\omega_k+\omega)}\right]
    \right\}\,,
\label{DstN}
\end{eqnarray}
where 
\begin{equation}
 \rho(k) =  {\omega_\mathrm{MIT}\over\omega_\mathrm{MIT}-1}\,
 {j_1(kR)\over\sqrt{2\pi}\,f_\pi\, R^3}\;.
\end{equation}
and $Z_N$ and $Z_\Delta$ are the usual 
wave-function-renormalization constants \cite{bg:thomas}.

The strong transition
form-factor $G_{\pi\mathrm{N}\Delta}(Q^2)$ is:
\begin{equation}
 {G_{\pi\mathrm{N}\Delta}(Q^2)\over 2M_N}
= 
 {\omega_\mathrm{MIT}\over\omega_\mathrm{MIT}-1}\,
  {1\over2f_\pi}\,
  {j_1(kR)\over kR}\,
\langle\Delta||\sum\sigma\,\tau||N\rangle
{2M_\Delta\over M_\Delta + M_N}\,.
\label{GpiNDcbm}
\end{equation}

Similarly as in (\ref{PCAC}) we can now explicitly show 
that PCAC is fulfilled provided the off-diagonal GT relation
holds in the model.
Since the Lagrangian is invariant under the chiral transformation 
both relation should  hold for the exact solution, 
but this is of course not obvious for the approximate solution.
In the model it is straightforward to evaluate the pertinent
quantities at $Q^2=-m_\pi^2$.
We prefer to give here the expressions at $Q^2=0$ which
take much simpler forms, e.g.:
\begin{equation}
C_5^A(0)
  = {1\over\sqrt{6}}
      \int\d r\,r^2\left[j_0(kr)
      \left(u^2 - {1\over3} v^2\right)
    -{4\over3}\,\,j_2(kr)v^2\right]
	\langle\Delta||\sum\sigma\,\tau||N\rangle 
{2M_\Delta\over M_\Delta + M_N}
\;,
\label{C5cbm}
\end{equation}
with $k=K_0$ (see (\ref{alphaW}).
 
The constant (\ref{C5cbm}) can be easily evaluated
for the degenerate N and $\Delta$ and neglecting
pion corrections to $\langle\Delta||\sum\sigma\,\tau||N\rangle$:
\begin{equation}
C_5^A(0)
  = {1\over\sqrt{6}}\,{3g_A^\circ\over5}\, 2\sqrt{2}
  = 0.755\;,
\label{C5est}
\end{equation}
where $g_A^\circ=1.09$ is the value of the nucleon $g_A$  
in the MIT bag model.
Clearly, (\ref{C5est}) strongly underestimates the experimental 
value (\ref{gAD}).
In the same limit, the strong coupling is
\begin{equation}
g_{\pi\mathrm{N}\Delta} \equiv 
  G_{\pi\mathrm{N}\Delta}(0)\,{m_\pi\over 2M_N}
 = \sqrt{72\over25}\,g^\circ_{\pi\mathrm{NN}} 
 =  1.39\;.
\label{GpiNDest}
\end{equation}
Here $g^\circ_{\pi\mathrm{NN}} = 0.82$ is the CBM value
without pion correction.
Again, (\ref{GpiNDest}) strongly underestimates 
the experimental value of 2.2,
though the off-diagonal Goldberger-Treiman relation
is exactly fulfilled in this approximation.

In \cite{bg:letter} we show that the pion corrections 
improve the results in particular the ratio of the strong 
$g_{\pi\mathrm{N}\Delta}$ and $g_{\pi\mathrm{NN}}$
coupling constants but the value of $C_5(0)$ remains
far below the experimental value.
A possible solution, described and discussed in 
\cite{bg:letter,bg:FB18,bg:miniBled} is to include 
the contribution of the $\sigma$-meson which enters 
the expression for the axial current (\ref{Anp}) 
and considerably increases the value of $C_5^A$.

\vspace{12pt}

This work was supported by FCT (POCTI/FEDER), Lisbon, and by
The Ministry of Science and Education of Slovenia.

\end{document}